\documentclass{appolb}
\usepackage{epsfig}

\newcommand{\leqsim}{\,\raisebox{-0.6ex}{$\buildrel < \over \sim$}\,}
\newcommand{\geqsim}{\,\raisebox{-0.6ex}{$\buildrel > \over \sim$}\,}
\def\ds{\displaystyle}
\def\etal{\mbox{\it et al.\ }}
\def\arxiv#1#2{[hep-#1/#2]}
\def\Journal#1#2#3#4{{#1} {\bf #2} {#3} ({#4})}

\def\PLB{{\it Phys. Lett.}  B}

\def\PRL{\it Phys. Rev. Lett.}
\def\PRD{{\it Phys. Rev.} D}

\def\ZPC{{\it Z. Phys.} C}
\def\EPJC{{\it Eur. Phys. J.} C}
\def\JPG{{\it J. Phys.} G.}

\begin{document}
\title{The rise and fall of $F_2$ at low $x$}
%
\author{A M Cooper-Sarkar \&  R.C.E Devenish
\address{Department of Physics, University of Oxford,
Denys Wilkinson Bldg, Keble Rd, Oxford OX1 3RH, UK}}
\maketitle
\begin{abstract}
A short personal account is given of the impact of HERA data and the 
influence of Jan Kwiecinski on low $x$ physics.
\end{abstract}

\section{Introduction}

\begin{figure}[tbp]
\vspace*{13pt}
\centerline{\psfig{figure=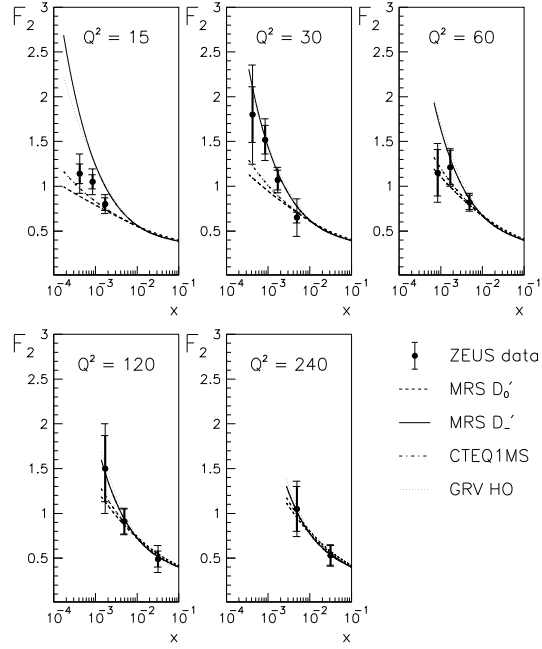,height=.5\textheight}} 
\caption{The first ZEUS $F_2$ data from HERA, published 1993 \cite{z93-f2}. }
\label{fig:f2-z93}
\end{figure}

\begin{figure}[tbp]
\vspace*{13pt}
\centerline{\psfig{figure=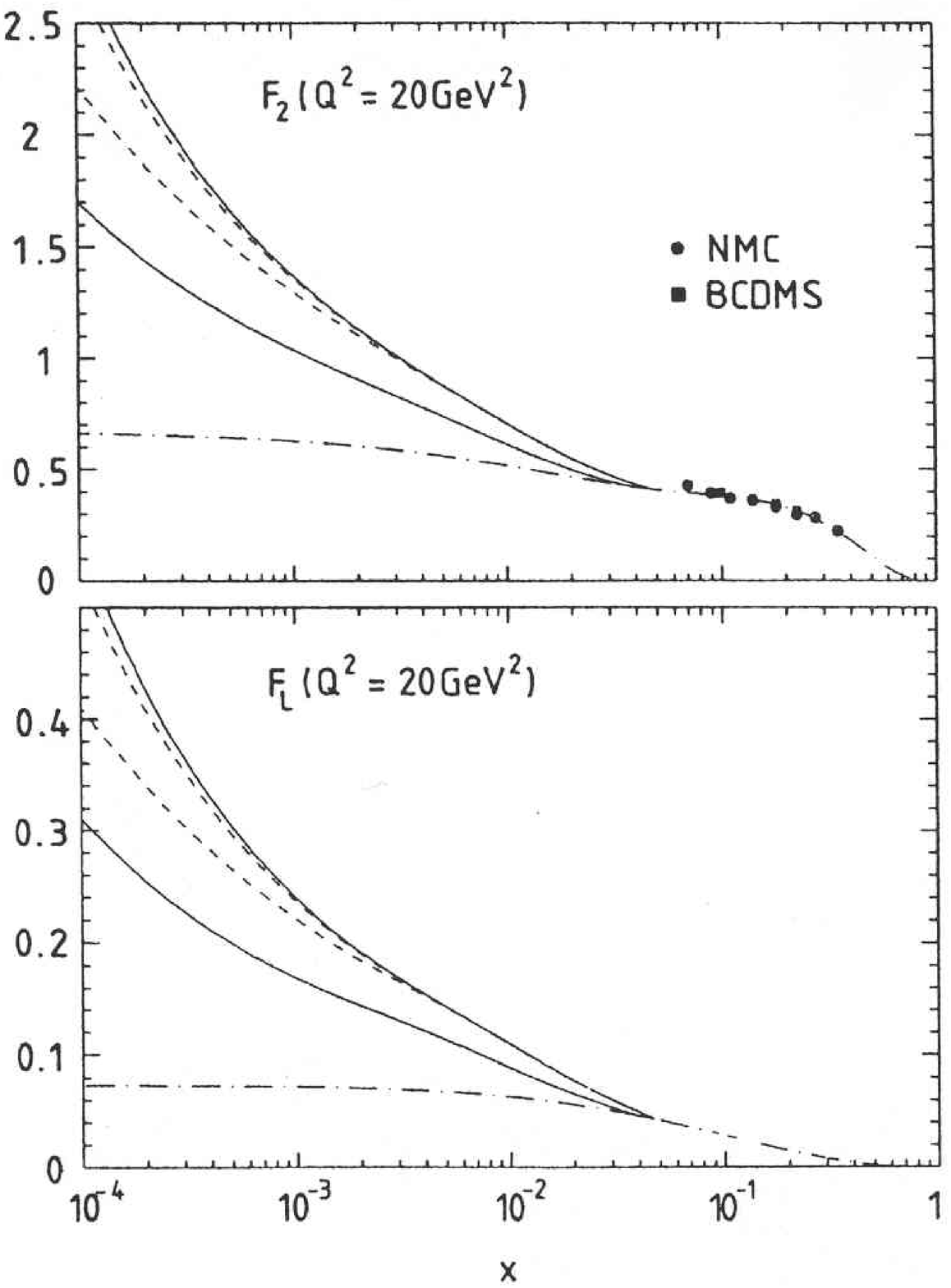,height=.5\textheight}} 
\caption{Calculations from AKMS \cite{akms} of the behaviour of $F_2$ and $F_L$
at low $x$ based on the BFKL equation. The difference between the upper
and lower continuous curves is in the cut-off imposed to control 
diffusion in transverse momentum ($k^2_0=1\,,2\,$GeV$^2$ for the upper, lower
curves respectively. The dashed curves show the effect of including shadowing
effects with a proton radius of $R=5,\,2\,$GeV$^{-1}$. The almost flat
dash-dotted curves are the contributions excluding the BFKL effects.}
\label{fig:akms-f2}
\end{figure}

That HERA data was to change our understanding of pQCD was announced at the 
Durham Phenomenology Workshop on HERA Physics in 1993 (the first of three
devoted to HERA Physics, all with a large input from JK). Albert De Roeck
presented the first preliminary H1 data from HERA on $F_2$ measurements at 
$x<0.01$, below the fixed target region. Within large errors that data 
showed a rising $F_2$ as $x$ decreased -- the quality of the data may be 
judged from Fig~\ref{fig:f2-z93} showing comparable ZEUS data.
The result was the major talking point of the 1993 HERA workshop and 
discussion has continued almost unabated since then.
One gets a sense of why the result was surprising by comparing 
Fig~\ref{fig:f2-z93} with Fig~\ref{fig:akms-f2}. The latter plot shows a
range of predictions from a paper by Askew, Martin, Kwiecinski \& Sutton
(AKMS) \cite{akms} with $F_2$ data from the NMC and BCDMS experiments for
$x>0.01$. The least model-dependent extrapolation from the measured 
data would appear give a `flat' $F_2$ as $x\to 0$ as shown by the 
dash-dotted curve, but what do the rising curves represent? 

\section{$F_2$ in pQCD}

A rising $F_2$ at small-$x$ is predicted by pQCD, but the predictions
do not give a scale in either $x$ or $Q^2$ at which one might expect
to see such effects.

In the DGLAP\footnote{Dokshitzer, Gribov, Lipatov, Altarelli \& Parisi,
collinear factorization.} formalism at leading order the gluon splitting 
functions
\begin{equation}
\ P_{gq} \to \frac{4}{3z},\ P_{gg} \to \frac{6}{z}
\label{eq:smallx}
\end{equation}
\noindent
are singular as $z\to 0$. Thus the gluon distribution will
become large as $x \to 0$, and its contribution to the evolution of the
parton distribution becomes dominant. In particular the gluon
will `drive'  the quark singlet distribution, and hence the structure 
function $F_2$, to become large as well, the rise increasing in
steepness as $Q^2$ increases. Quantitatively,
\begin{equation}
 \frac{dg(x,Q^2)}{d \ln Q^2} \simeq \frac{\alpha_s(Q^2)}{2\pi} \int^1_x
\frac{dy}{y} \frac{6}{z} g(y,Q^2)
\label{eq:glev}
\end{equation}
\noindent
may be solved subject to the nature of the boundary function $xg(x,Q^2_0)$.
Inputting a non-singular gluon at $Q^2_0$, the solution is~\cite{deRujula,das} 
\begin{equation}
 xg(x,Q^2) \simeq \exp\left(2 \left[\xi (Q^2_0,Q^2)\ln\frac{1}{x}\right]
^{\frac{1}{2}}\right)
\label{eq:DLLAg}
\end{equation}
\noindent
where
\begin{equation}
 \xi(Q^2_0,Q^2) = \int^{Q^2}_{Q^2_0} \frac{dq^2}{q^2} \frac{3\alpha_s(Q^2)}
{\pi}
\label{eq:DLLAxi}
\end{equation}
\noindent
Given a long enough evolution length from $Q^2_0$ to $Q^2$, this will
generate a steeply rising gluon distribution at small $x$, starting from 
a flattish behaviour of $xg(x,Q^2)$ at $Q^2 = Q^2_0$. 

Over the $x,Q^2$ range of HERA data this solution mimics a power 
behaviour, $xg(x,Q^2) \sim x^{-\lambda_g}$, with 
\begin{equation}
\lambda_g = \left(\frac{12}{\beta_0}
\frac{ \ln(t/t_0)}{\ln(1/x)}\right)^{\frac{1}{2}}
\end{equation}
\noindent
where $t=\ln(Q^2/\Lambda^2)$, $t_0=\ln(Q_0^2/\Lambda^2)$.
This steep behaviour of the gluon generates a similarly
steep behaviour of $F_2$ at small $x$, $F_2 \sim x^{-\lambda}$, where
$\lambda = \lambda_g -\epsilon$.

So why was the observed rise of $F_2$ unexpected?  Because, before the advent
of HERA data, the starting  scale for perturbative evolution had always been 
taken to be $Q^2_0 \geqsim 4\,$GeV$^2$, to be sure that a perturbative 
calculation would be valid. The evolution length from $Q^2$ of 4 to 
$15\,$GeV$^2$ is not large enough to generate a steep slope from a flat input.
\begin{figure}[tbp]
\vspace*{13pt}
\centerline{\psfig{figure=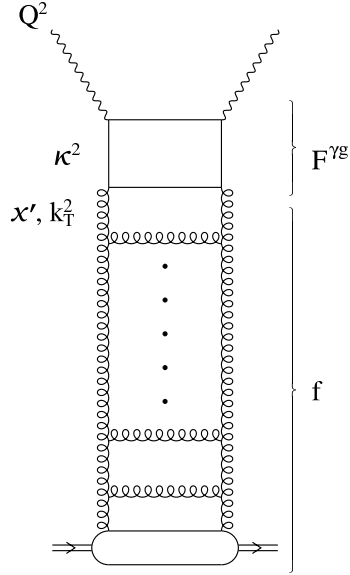,height=.4\textheight}} 
\caption{The BFKL gluon ladder diagram for deep inelastic scattering with a
quark box at the top connecting to the $\gamma^*$ and the proton at the
bottom.}
\label{fig:ladder}
\end{figure}

However Jan Kwiecinski together with Martin, Roberts and Stirling
\cite{kmrs} had already been considering
alternatives to `conventional' DGLAP with flat input distribution. 
The KMRSB$_0$ and KMRSB$_-$ parton distribution functions of 1990 \cite{kmrs}
were both compatible with the data: KMRSB$_0$ had a conventional 
flat gluon distribution input at $Q^2_0 = 4\,$GeV$^2$; 
whereas KMRSB$_-$ had a singular gluon distribution $x^{-0.5}$, even at 
low $Q^2(=Q^2_0$).

 Why should one consider a steep input gluon?. Because at small $x$ terms in
$\ln(1/x)$ are becoming large and the conventional {\it leading $\ln Q^2$} 
summation of the DGLAP equations does not account for this. It may also be 
necessary to sum {\it leading $\ln(1/x)$} terms. Such
a summation is performed by the BFKL\footnote{Balitsky, Fadin, Kuraev \&
Lipatov, $k_T$ factorization.} equation. To leading order in
$\ln(1/x)$ with fixed $\alpha_s$, this predicts a steep power law behaviour
\begin{equation}
 xg(x,Q^2) \sim f(Q^2)\ x^{-\lambda_g}
\label{eq:BFKLsol}
\end{equation}
\noindent 
where 
\begin{equation}
 \lambda_g = \frac{3\alpha_s}{\pi} 4\ln2 \simeq 0.5
\label{eq:BFKLlam}
\end{equation}
\noindent 
(for $\alpha_s \simeq 0.2$, as appropriate for $Q^2 \sim 4\,$GeV$^2$).

These ideas were taken further in the paper by AKMS, see 
Fig.~\ref{fig:akms-f2}, in which BFKL evolution was incorporated directly
into the calculation of the parton densities. The rising curves show
different calculations of $F_2$ and $F_L$ from the BFKL equation, the 
dashed curves show the
effect of including some damping by shadowing which will be discussed
later. The dash-dotted curve is the prediction for the structure functions
without BFKL effects. 
 
Thus the observation of a steep behaviour in $F_2$ at a relatively low $Q^2$
was seized upon by those who saw this as evidence for BFKL. On the other
hand it was quickly realised that by lowering the starting scale in
the DGLAP approach to $Q^2_0\approx 1\,$GeV$^2$ one could also fit the data
without a singular gluon input distribution. This later approach owed
much to the work of Gl\"uck, Reya and Vogt (GRV) \cite{grv1}. Higher
statistics data taken during 1993 showed that an effective power 
$\lambda_g\approx 0.5$ was actually too steep.
\begin{figure}[tbp]
\vspace*{13pt}
\centerline{\psfig{figure=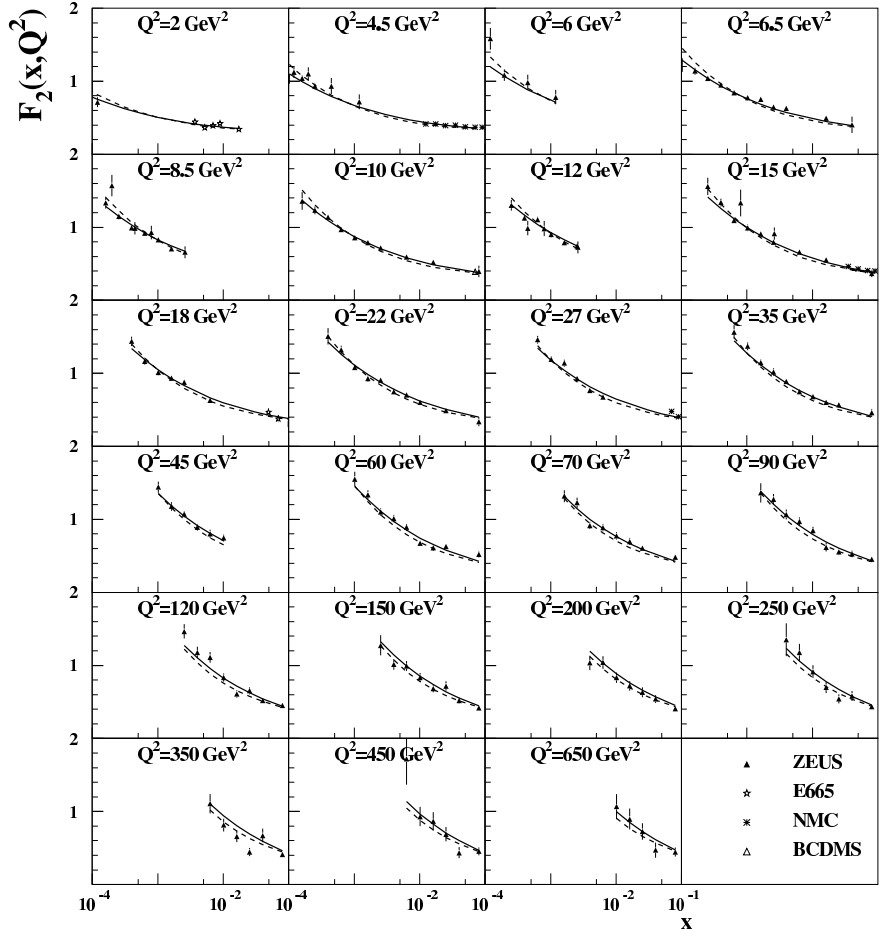,height=.65\textheight}} 
\caption{The combined BFKL-DGLAP model of KMS fit to $F_2$ data from
ZEUS, E665, BCDMS \& NMC. From \cite{kms}.}
\label{fig:kms-fg3}
\end{figure}

Kwiecinski was involved with many of the approaches to find a credible
phenomenology of the BFKL equation. One of the most significant of these is 
the need for a `kinematic constraint' to control the 
diffusion of transverse momentum down the `gluon ladder' in the BFKL
approach (see Fig.~\ref{fig:ladder}). 
\begin{figure}[tbp]
\vspace*{13pt}
\centerline{\psfig{figure=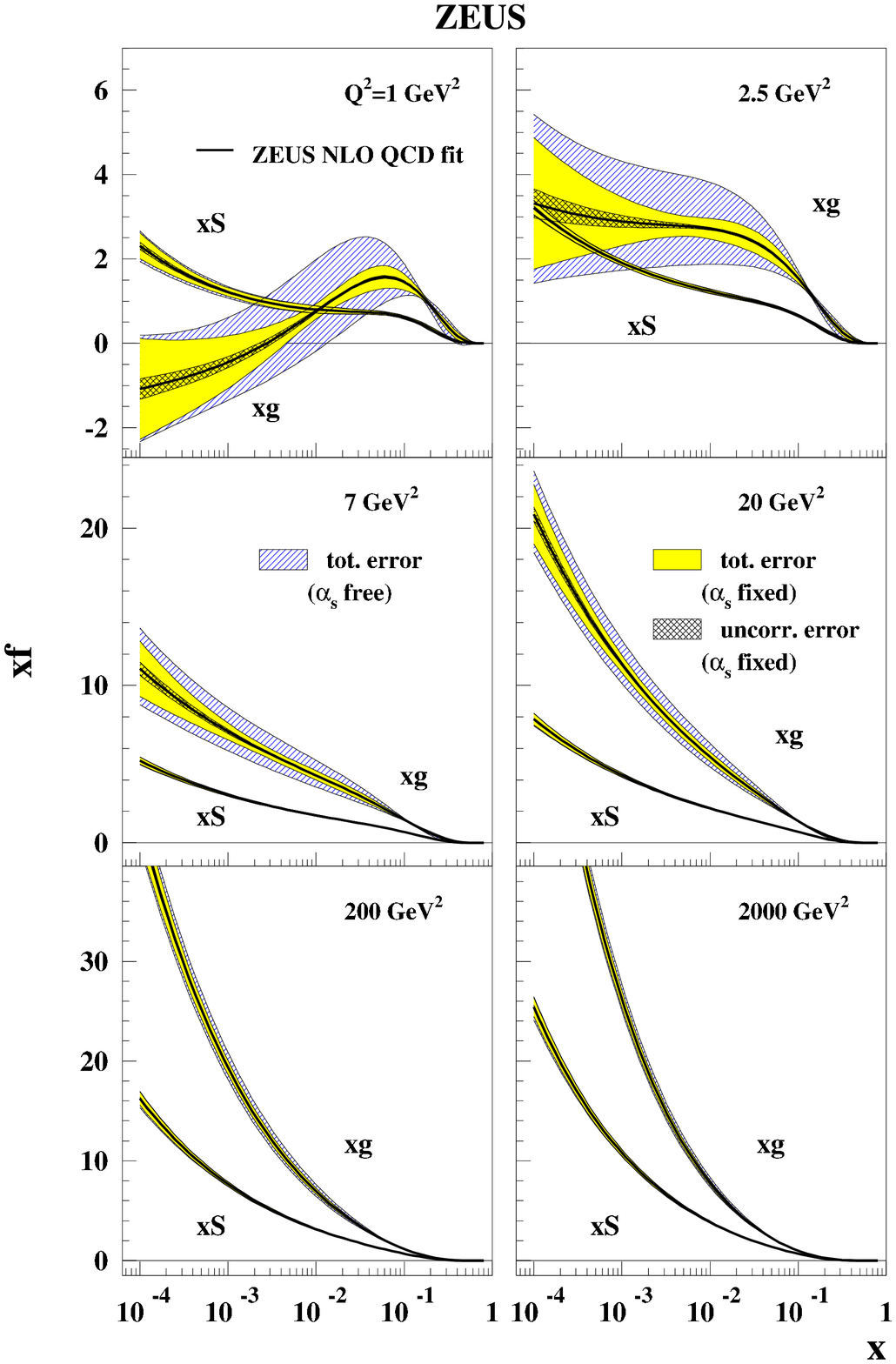,height=.6\textheight}} 
\caption{Gluon and sea quark momentum distributions compared. From a NLO
QCD fit by ZEUS using the full HERA-I data sample \cite{z02qcd}.}
\label{fig:glusea}
\end{figure}
Unlike the DGLAP summation in which
the $k_T$ of the ladder are strictly ordered, those in the BFKL summation are
not and can diffuse downwards into the non-perturbative region. 
Consider a link in the gluon chain where the longitudinal momentum
fraction decreases from $x/z$ to $x$ and the transverse momentum $k_T'$
changes to $k_T$, with emission of a gluon of transverse momentum $q_T$. 
One requires 
\begin{equation}
k_T^2/z > q_T^2 
\label{eq:const}
\end{equation}
\noindent
in order that the virtuality of exchanged gluons is controlled by their 
transverse momenta. This implies that $k_T^2/z > k_T'^2$, for any given 
value of $k_T$. If one considers the effect of this constraint on the 
solutions for the BFKL equation one finds that it modifies the  asymptotic 
solution $x^{-\lambda_g}$, such that $\lambda_g$ is reduced from
$\lambda_g \sim 0.5$ to $\lambda_g \sim 0.3$. These ideas were built into
an ambitious approach by Kwiecinski, Martin and Stasto \cite{kms} to
combine the BFKL and DGLAP approaches in a single model to be applied to
the HERA and fixed target data. The key idea was to use the BFKL kernel,
with the kinematic constraint, applied to the unintegrated gluon density
(i.e. differential in $k_T$ as well as $x$) to give an input gluon density
to DGLAP evolution. This gives rise to a pair of coupled equations for
the gluon and $q\bar{q}$ sea which can be solved numerically. The small
valence quark contribution was taken from the leading order GRV set
\cite{grv2} to give a complete model for $F_2$, $F_2^{charm}$ and $F_L$.
Only two 
parameters were needed to describe $xg(x,k^2_0)$ at the starting scale.
A very good fit to the HERA 1994 $F_2$ data together with data from E665,
BCDMS and NMC was obtained, as shown in Fig.~\ref{fig:kms-fg3}.  

During the period $1994-1996$ the conventional DGLAP NLO QCD global fits were
also refined and gave very good $\chi^2$ fits to the HERA data,
starting with a conventional flat gluon input, provided that a low 
starting scale $Q^2_0 = 1\,$GeV$^2$ was used.
In 1996 the HERA data from the 1994 run were published, showing that
that the steep slope of $F_2$ extended as far down in $Q^2$ as 
$Q^2 \sim 1.5\,$GeV$^2$. The precision of the data had now increased 
sufficiently that the low-$x$ behaviour of the sea and the gluon distributions
could be fitted separately. These results led to a new kind of surprise since
it became clear that the flat input which had been used for both the sea and 
the gluon ($\lambda_S = \lambda_g \sim 0$) was actually a compromise between
a sea distribution which remains steep, $\lambda_S \sim 0.2$ down to 
$Q^2 \sim 1~$GeV$^2$, and a gluon distribution which is becoming valence-like, 
$\lambda_g < 0$ at low $Q^2$. Fig.~\ref{fig:glusea} shows a recent result from
a ZEUS global fit using the complete HERA-I data sample \cite{z02qcd}. 
This behaviour contradicts the
original argument that the  steep behaviour of the gluon distribution is 
driving the steep behaviour of the sea which is measured in $F_2$.
\begin{figure}[tbp]
\vspace*{13pt}
\centerline{\psfig{figure=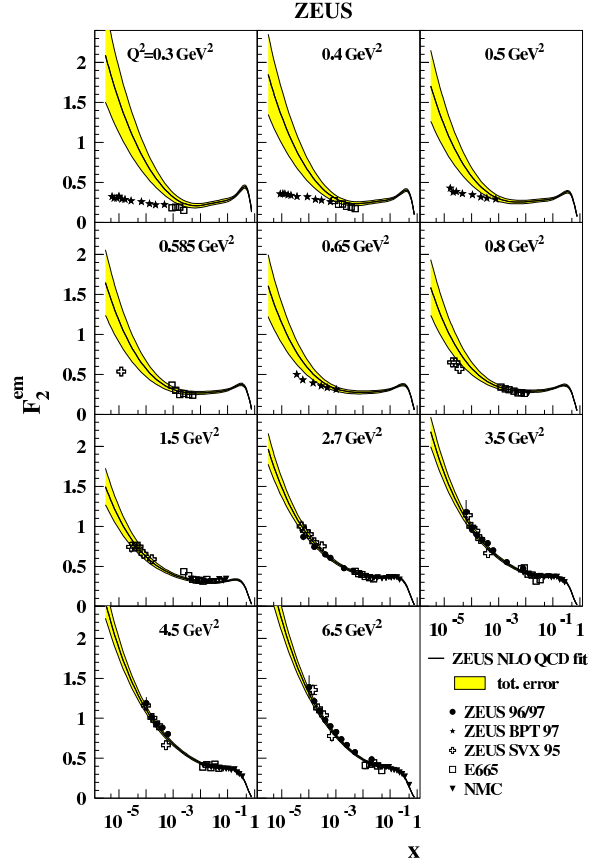,height=.6\textheight}} 
\caption{ZEUS $F_2$ data at low $Q^2$ compared to the ZEUS-S pQCD fit
\cite{z02qcd}.}
\label{fig:lowq2}
\end{figure}

These conclusion were strengthened when the data from the 1996/7 
runs were published in 2000/2001. It became possible to make detailed analyses
of the experimental uncertainties on the parton distributions and thus put
limits on the kinematic range of applicability of the DGLAP formalism. 
For  $Q^2 \leqsim 1$GeV$^2$ the gluon
distribution becomes negative at small $x$, see Fig.~\ref{fig:glusea} 
and the fit can no longer describe the data, see Fig.~\ref{fig:lowq2}.
Furthermore, even though 
the DGLAP formalism works for $Q^2 \geqsim 1~$GeV$^2$, 
doubts remain about its applicability when 
$\lambda_g < \lambda_S$. With only one observable measured with high
precision and all models depending on parameters fit to data, one has
sufficient flexibility to get good descriptions in both the pure
DGLAP and `BFKL enhanced' approaches.
The striking rise in $F_2$ does begin to abate 
for $Q^2$ below $2\,$GeV$^2$ as Fig.~\ref{fig:lowq2} illustrates. 
At this point is useful to bring in
data on the slopes or derivatives of $F_2$.

\section{$F_2$ slopes} 

\begin{figure}[tbp]
\vspace*{13pt}
\centerline{\psfig{figure=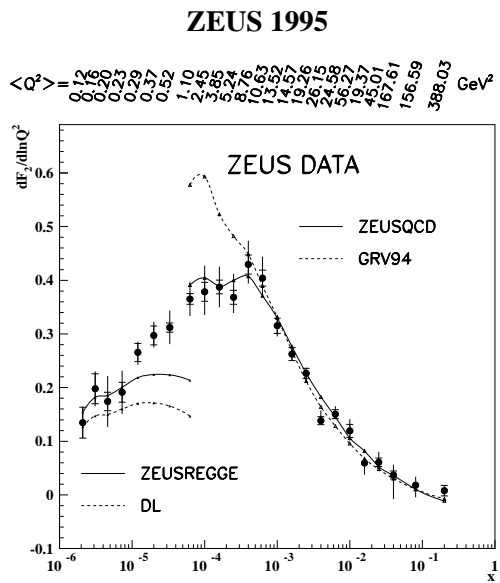,height=.4\textheight}} 
\caption{The slope  $\ds {\partial F_2\over \partial \ln Q^2}$ as a
function of $x$ from ZEUS 1994/5 HERA data. From \cite{Zphenom}.}
\label{fig:df2dlnq2}
\end{figure}
The behaviour of $F_2$ may be characterized in terms of two slopes,
with respect to $\ln Q^2$ and $\ln(1/x)$ respectively. The latter
has already been used in the discussion of the gluon density at low $x$,
here $\ds \lambda(Q^2) = {\partial F_2\over \partial \ln(1/x)}$ or
$\ds F_2(x,Q^2)\sim x^{-\lambda(Q^2)}$. At low $x$ the form of the DGLAP
equations is such that at LO one has (very roughly) 
\[
F_2(x,Q^2)\sim x\Sigma(x,Q^2),~~~{\rm and}~~~
{\partial F_2(x,Q^2)\over \partial \ln Q^2}\sim  P_{qg}\otimes xg(x,Q^2).
\]
 
Thus the behaviour of $\partial F_2/\partial\ln Q^2$ may come from 
either the gluon density
or the splitting function or both. If it could be shown that any
unconventional behaviour was attributable to $P_{qg}$ then this would
indicate the need for an alternative evolution equation. 
In any case it is well worthwhile the experimental groups calculating the two
slopes with a proper treatment of systematic errors. 
$\ds {\partial F_2\over \partial \ln Q^2}$ -- the `Caldwell plot'
\footnote{A preliminary version of this plot was first shown by Allen Caldwell
at the 1997 DESY Theory Workshop.} as a function of $x$ is shown in 
Fig.~\ref{fig:df2dlnq2}. This is another result from HERA that caused a lot 
of interest and discussion! The plot shows the slopes calculated from the
$F_2$ shown in Fig.~\ref{fig:kms-fg3}. An important detail to note is that
because of the strong correlation between $x$ and $Q^2$ inherent in deep
inelastic kinematics, the slopes are not evaluated at a fixed values of
$Q^2$, but rather at the mean values shown along the top of the plot for
each data point. The plot shows that the rate of rise of $F_2$ as a
function of $Q^2$ is starting to abate as $x<10^{-3}$ and $Q^2<8\,$GeV$^2$.
To understand what the plot might indicate and why one is also very
interested in where and how $F_2$ stops rising at low $x$
it is helpful to look at the HERA data from another point of view. 

\section{$\sigma(\gamma^*p)$ and saturation}

The strong rise of $F_2$ at low $x$ was unexpected for another reason.
At small $x$, $F_2$ is related to the cross-section for $\gamma^*p$
scattering by
\begin{equation}
\sigma^{\gamma^*p}(W^2,Q^2) \approx \frac{4\pi^2\alpha}{Q^2} F_2(x,Q^2)
,~~~{\rm with}~~~W^2\approx Q^2/x,
\label{eq:sigf2}
\end{equation}
where $W$ is the $\gamma^*p$ centre of mass energy. This relation implies
that a rise of $F_2$ at small-$x$ corresponds to a rising $\gamma^* p$ 
cross-section with $W^2$. The relationship above suggests
another approach to understanding the behaviour of $F_2$ at low $x$ and
that is the framework of Regge theory used to explain the high energy
behaviour of hadronic total cross-sections. Regge theory predicts that 
the high energy behaviour of hadronic scattering amplitudes is 
$\ds {\rm Im} A(ab \to cd) \sim \sum_i \beta_is^{\alpha_i}$, where
$\alpha_i$ is the intercept of a Regge trajectory which has the 
right quantum numbers for an exchange in the crossed channel 
$a\bar{c} \to \bar{b}d$. Using the optical theorem, a total cross-section
will vary as $\ds \sigma^{tot}(ab) \sim \sum_i \beta_is^{\alpha_i -1}$. 
For the corresponding forward elastic scattering amplitude, the leading  Regge 
trajectory has the quantum numbers of the vacuum and is known as the Pomeron,
for which the value $\alpha_P \approx 1.08$ has been determined from 
hadron-hadron data. This prediction also describes the high energy behaviour
of photoproduction cross-section measurements successfully. 
The model was extended to describe virtual-photon proton scattering by 
Donnachie and Landshoff~\cite{dl1} who assumed that the $Q^2$ 
dependence would reside in the residue functions $\beta_i(Q^2)$ only
and that the intercepts, $\alpha_i$ would be independent of $Q^2$.
Through Eq.~\ref{eq:sigf2} this approach then predicts a flattish input 
for the gluon and sea PDFs, since $\sigma(\gamma^* p) \sim s^{\alpha_i-1}$ 
implies $F_2 \sim x^{1-\alpha_i} = x^{-0.08}$.
\begin{figure}[tbp]
\vspace*{13pt}
\centerline{\psfig{figure=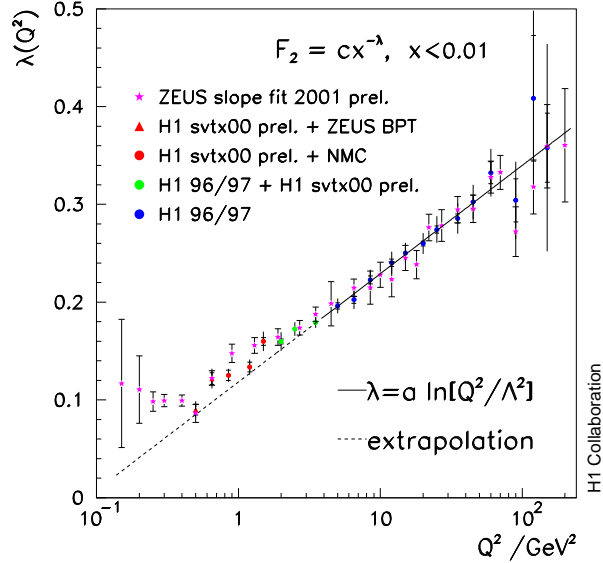,height=.4\textheight}} 
\caption{The effective slope of $F_2$ at low $x$, 
$\ds F_2\sim x^{-\lambda(Q^2)}$. From \cite{gayler}.}
\label{fig:lambdaeff}
\end{figure}
While the DL approach successfully described the pre-HERA low $x$ data for 
$Q^2$ values up to about $10\,$GeV$^2$, it cannot describe the steeply 
rising $F_2$ data measured at HERA. The problem is summarised graphically
in Fig.~\ref{fig:lambdaeff} which shows 
$\ds \lambda (Q^2) = {\partial\ln F_2\over \partial\ln (1/x)}$ or  
$\ds F_2 \sim x^{-\lambda(Q^2)}$, at low $x$. For $Q^2$ values less
than $1\,$GeV$^2$ or so, the value of $\lambda$ is consistent with that
from hadronic Regge theory, whereas for $Q^2>1\,$GeV$^2$ the slope
rises steadily to reach a value greater than $0.3$ by 
$Q^2\approx 100\,$GeV$^2$. This larger value of $\lambda$ is not so far
from that expected using BFKL ideas. Indeed the
BFKL equation can be viewed as a method for calculating the hard or
perturbative Pomeron trajectory -- in contrast to the soft or 
non-perturbative Pomeron of hadronic physics with intercept around
$1.08$. Donnachie and Landshoff \cite{dl2} have extended their Regge model 
by the addition of a hard Pomeron with intercept $1.44$, which allows them
to describe the low $x$ HERA data up to $Q^2$ values of a few hundred
GeV$^2$. Theirs is one of many attempts, using ideas from Regge theory,
to model the `transition region' from real photoproduction through to deep 
inelastic scattering with $Q^2>>1\,$GeV$^2$. 

Either considering the rise of $F_2$ as $x\to 0$ or the steep energy
dependence of $\sigma(\gamma^*p)$ at large $Q^2$, such behaviour will 
eventually violate the Froissart bound. If the origin of the rise arises
from a high gluon density then this problem may be avoided because the 
gluons can shadow each other from the $\gamma^*$. At even higher
densities the gluons will `recombine' via the process $gg\to g$ -- the inverse 
of gluon splitting -- and the gluon density and hence $F_2$ will saturate.
These ideas have been formalized by Gribov, Levin \& Ryskin by the
addition of a non-linear term to the equation for gluon evolution
\begin{equation}
\frac{d^2 xg(x,Q^2)}{d lnQ^2 dln(1/x)} = \frac{3\alpha_s}{\pi} xg(x,Q^2) -
\frac{81\alpha_s^2}{16 Q^2 R^2}\left[xg(x,Q^2)\right]^2 
\end{equation}
When $xg(x,Q^2) \sim \pi Q^2 R^2 /\alpha_s(Q^2)$ the non-linear
term cancels the linear term and evolution stops, this is saturation. 
Shadowing effects were included in the calculations by AKMS shown in 
Fig.~\ref{fig:akms-f2} and in the KMRS parton densities.
 
\begin{figure}[tbp]
\vspace*{13pt}
\centerline{\psfig{figure=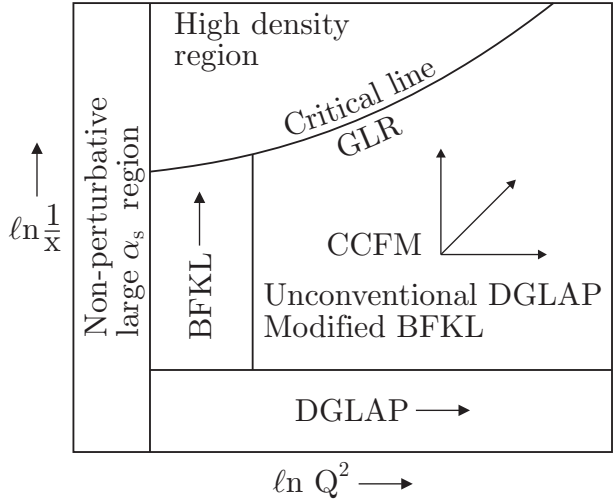,height=.35\textheight}} 
\caption{Approaches to physics at low-$x$. Courtesy of A D Martin}
\label{fig:admmap}
\end{figure}

The various evolution equations applicable across the $x,Q^2$ 
plane are summarised schematically 
in Fig.~\ref{fig:admmap}.\footnote{An early version of this figure
was shown by JK in his plenary talk on low $x$ QCD at the 1993 Durham Workshop
\cite{jkdur93}.}
Note the appearance of the `critical line' above which the gluon density
is high enough for non-linear effects to be important. In this region and
for large enough $Q^2$ one has the possibility that $\alpha_s$ will be 
small enough for weak-coupling but non-perturbative methods to be applicable, 
of which the GLR equation could be a first approximation. 
Again no scales are given
and another of many hotly argued questions raised by the HERA data --
for example in the context of the Caldwell plot (Fig.~\ref{fig:df2dlnq2}) 
-- is does HERA have the reach to see saturation effects and have they been
seen?

Dipole models have proved to be a very fruitful approach in exploring this
question and modelling the behaviour of $F_2$ or $\sigma(\gamma^*p)$
through the transition region towards $Q^2=0$. The idea is that the 
virtual photon splits into a $q \bar{q}$ pair\footnote{As for example shown
by the quark box at the top of Fig.~\ref{fig:ladder}.} 
(a colour dipole) with a transverse size $r\sim 1/Q$, the splitting 
occurring a `long' time $\sim 1/(mx)$ before the dipole interacts with the 
proton. The dipole then scatters coherently from the proton in a time 
which is short in comparison to $\sim 1/(mx)$. 
The $\gamma^*\to q \bar{q}$ 
process is described by QED and the strong interaction physics 
is then contained in the cross-section, $\hat{\sigma}$,  for the 
dipole-proton interaction. 

Many authors have worked on these models in
the context of DIS processes both inclusive and for diffraction. Here
the discussion will focus on the model of Golec-Biernat and W\"usthoff (GBW) 
\cite{gbw}. The original GBW model provides a simple
parameterization of $\hat{\sigma}$ with explicit saturation, in the sense
that $\hat{\sigma}\to \sigma_0$ (a constant) as $r$ becomes large, but 
in a such a way that the approach to saturation is $x$ dependent, controlled
by the `saturation radius', $R_0$. Explicitly the dipole cross-section is
given by
\begin{equation}
\hat{\sigma}(x,r^2)=\sigma_0 (1-\exp(-\hat{r}^2)),~~~\hat{r}={r\over 2R_0(x)}
\label{eq:gbw-g}
\end{equation}
where $\sigma_0$ is a constant, and
\begin{equation}
R_0(x)^2={1\over Q_0}\left({x\over x_0}\right)^{\lambda},
\end{equation}
$R_0(x)^2$ is understood to be inversely proportional to the 
gluon density ($\lambda \sim \lambda_g$), so that $R_0$ is a measure of the
transverse separation of the gluons in the target. Thus when the dipole 
separation is large compared to the gluon separation  (small $Q^2$ and 
small $x$) the dipole cross-section saturates, $\hat{\sigma} \sim \sigma_0$, 
and $\sigma(\gamma^*p)$ also tends to a constant, giving 
$F_2 \propto 1/Q^2$ from Eq.~\ref{eq:sigf2}. When the dipole separation is 
small compared to the gluon separation (high $Q^2$ and large $x$),
$\hat{\sigma} \sim \sigma_0/(Q^2 R_0^2)$ 
and $\sigma(\gamma^*p)$ varies as $1/Q^2$, so that $F_2$ exhibits Bjorken 
scaling. Because of the $x$ dependence of $R_0$, $\hat{\sigma}$ saturates
for smaller dipoles sizes as $x$ decreases. 

\begin{figure}[tbp]
\vspace*{13pt}
\centerline{\psfig{figure=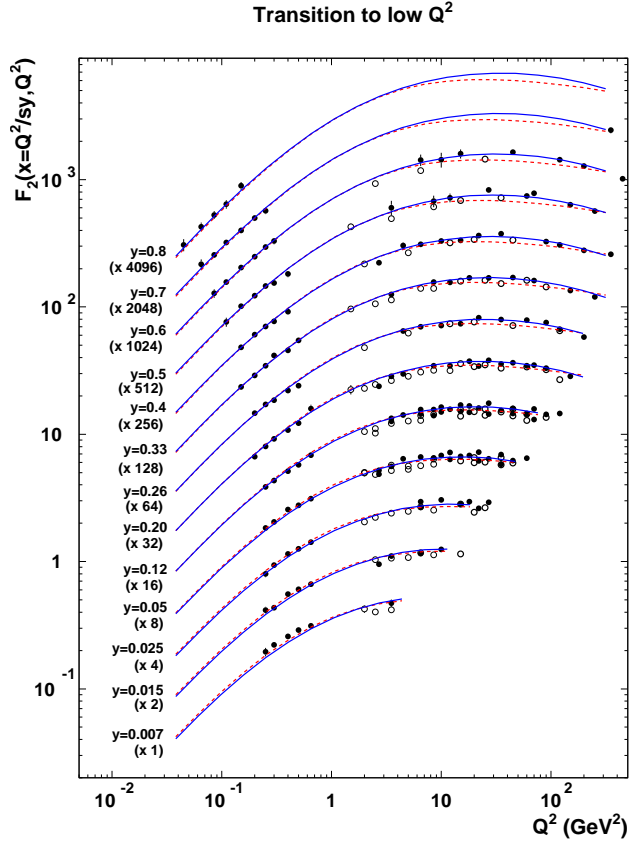,height=.6\textheight}} 
\caption{The GBW dipole model fit to $F_2$ data through the transition
region to photoproduction. The data are plotted versus $Q^2$ in bins
of $y$. The dashed curves show the original model, the full curve the
improved model with DGLAP evolution. From \cite{gbbk}.}
\label{fig:gbw-f2}
\end{figure}

Looking at Fig.~\ref{fig:gbw-f2},
which shows $F_2$ as a function of $Q^2$ at fixed $y=Q^2/sx$, one sees that 
the data do exhibit these features. There is a clear change in behaviour 
around $Q^2\approx 1\,$GeV$^2$, which might then be taken as a rough estimate 
of the saturation scale for HERA data.
However, one needs to be somewhat careful before jumping to the conclusion that
the GBW dipole model shows that saturation effects have
been seen at HERA. Given Eq.~\ref{eq:sigf2},
$F_2 \sim 1/Q^2$ at low $Q^2$ is required in any model since
the photoproduction cross-section is finite and slow,
logarithmic scale breaking at high $Q^2$ is required by pQCD. Thus the real 
challenge is not in describing the general
behaviour but in describing the exact shape of the data across the transition,
now that high precision data are available. It should also be noted that
there are other dipole models that fit the same data as successfully without
an explicit $x$ dependent saturation scale.

Although it gave a good description of $F_2$ at low $Q^2$, the original
GBW model did not include QCD evolution in $Q^2$ and this limited its
success at larger $Q^2$. This deficiency was corrected in a second version
\cite{gbbk} in which $\hat{\sigma}$ was related to the gluon density. Results
from both versions are shown in Fig.~\ref{fig:gbw-f2}.

\begin{figure}[tbp]
\vspace*{13pt}
\centerline{\psfig{figure=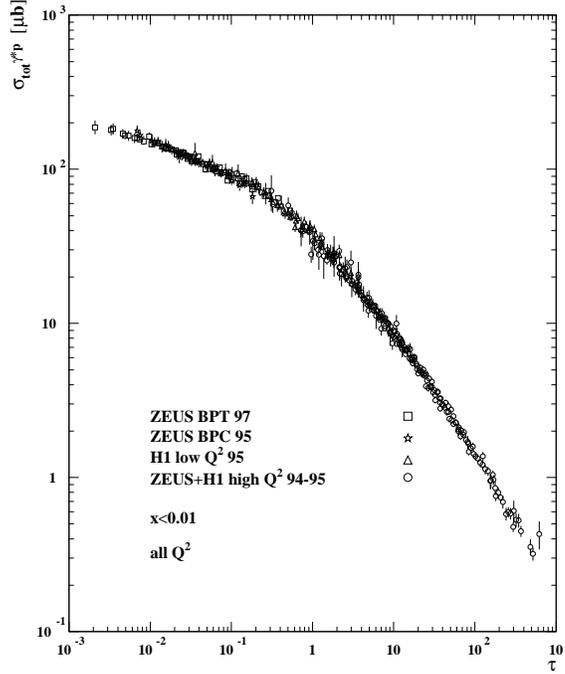,height=.5\textheight}} 
\caption{Geometrical scaling. Data on $\sigma(\gamma^*p)$ with
$x<0.01$ plotted versus the scaling variable $\tau=Q^2R^2_0(x)$.
From \cite{Sta01}.}
\label{fig:gscal}
\end{figure}
Another feature of saturation models which is illustrated in
the GBW model is the dependence of the dipole cross-section on a
scaled variable, here $r/R_0(x)$. Stasto, Golec-Biernat and Kwiecinski 
\cite{Sta01} have made the interesting observation that this leads to 
a new scaling 
property of $\sigma(\gamma^* p)$. At low $x$, $\sigma(\gamma^* p)$ should 
depend only on the dimensionless variable $\tau=Q^2R^2_0$. What is more, 
such a scaling property seems to be satisfied by HERA data. One expects
\[
\sigma(\gamma^* p)\sim \sigma_0~~(\tau~{\rm small}) \to
\sigma(\gamma^* p)\sim \sigma_0/\tau~~(\tau~{\rm large}).
\]
Fig.~\ref{fig:gscal} shows HERA data with $x<0.01$ as a function of
$\tau$. Not only do the data show this scaling very clearly, but also 
the expected change in behaviour indicated above seems to occur around 
values of $\tau\sim 1$. This new scaling -- {\it 
geometrical scaling} -- holds for $Q^2$ values up to about $400\,$GeV$^2$
but only for $x<0.01$. Again, though it is not a proof, the remarkable
demonstration of geometrical scaling by the low $x$ HERA data does show that
they have many of the attributes of a saturated system. 

\section{Summary}

The low $x$ data from HERA have produced many surprising and interesting
results that can be understood in the perturbative dynamics of a gluon
rich system, although details such as exactly how to include
BFKL effects are as yet undecided. Whether saturation effects have been
demonstrated at HERA is controversial, but whatever the answer to this 
question one cannot deny the importance of the measurements in 
stimulating activity in this area. Particularly encouraging are the
attempts to describe high density gluon dynamics in terms of semi-classical
models -- the so-called `colour glass condensate', which could be
important for understanding data from heavy ion collisions at RHIC and
the LHC.

Jan Kwiecinski has been active in most of these fields and his work can be
seen as an attempt to understand and set scales on the map shown in
Fig.~\ref{fig:admmap}. It has been a pleasure to profit from his insights
and untiring efforts to unravel the dynamic structure of the proton.

\end{document}